\documentclass[aps,pra,twocolumn,showpacs,groupedaddress]{revtex4-1}
\usepackage{graphicx}  
\usepackage{dcolumn}   
\usepackage{bm}        
\usepackage{amssymb}   
\usepackage{amsmath}
\usepackage{braket}
\usepackage[mathscr]{euscript}
\usepackage{color}
\usepackage{epsfig,color}
\usepackage{hyperref}
\begin{document}
\title{Transitionless quantum driving based wireless power transfer}
\author{Koushik Paul}
\email{p.koushik@iitg.ernet.in}
\author{Amarendra K Sarma}
\email{aksarma@iitg.ernet.in}
\affiliation{Department of Physics, Indian Institute of Technology Guwahati\\ Guwahati-781039, Assam, India}
\date{\today}
\begin{abstract}
Shortcut to adiabaticity (STA) techniques have the potential to drive a system beyond the adiabatic limits. Here, we present a robust and efficient method for wireless power transfer (WPT) between two coils based on the so-called transitionless quantum driving (TQD) algorithm. We show that it is possible to transfer power between the coils significantly fast compared to its adiabatic counterpart. The scheme is fairly robust against the variations in the coupling strength and the coupling distance between the coils. Also, the scheme is found to be reasonably immune to intrinsic losses in the coils.
\end{abstract}

\maketitle
\section{\label{sec1}Introduction}

Techniques for coherent manipulation of quantum states, specially for two-level and three-level quantum systems, are of great interest in modern day quantum optics and atomic physics. In this context, resonant excitation via pulsed radiation and adiabatic evolution by means of frequency sweep, with controlled chirp across resonance, are particularly well known. These two methods have been investigated theoretically and demonstrated experimentally by a wide range of implementations in past few decades \cite{1}. To reach at a desired target state, from a given one, using resonant excitation, a constant frequency $\pi$ pulse is required. One needs to have precise control over the pulse area which makes it sensitive to the fluctuations in various parameters. The adiabatic processes like rapid adiabatic passage (RAP), Stark chirped rapid adiabatic passage (SCRAP), stimulated Raman adiabatic passage (STIRAP) etc.\ do not require such precise control and are robust against parameter fluctuations \cite{2}. However, for adiabatic evolution, the system must evolve adiabatically with time making those processes relatively slow. Furthermore, no matter how robust the process is, it is necessary to drive a system as fast as possible in order to reduce the effects of definite decoherences present in the system. Efforts have been made to address those issues in recent years. Current developments in this direction are mainly based on the idea of accelerating the evolution beyond the adiabatic regime without violating the principles of state transfer. Few  shortcut to adiabaticity (STA) techniques, such as {\it transitionless quantum driving} (TQD) \cite{3}, {\it counter diabatic algorithm} (CDA) \cite{4} and {\it Lewis-Riesenfeld invariant} (LRI) approach \cite{5} are put forward to speed up the time evolution of quantum systems. In past few years, these methods have been explored rigorously across various branches of physics such as waveguide couplers \cite{6}, Bose-Einstein condensates \cite{7}, entangled state preparation \cite{8}, non-hermitian systems \cite{9} and so on. Owing to the ubiquitous nature of adiabatic processes, STA techniques span a broad range of applications. One possible potential use of these techniques may be in the so-called wireless power transfer technology.

In modern age, various wireless technologies play crucial role in our day-to-day life. Since the early days of electromagnetics, significant progress has been made in transferring information via wireless communication. In contrast, wireless power transfer (WPT) gained little progress in the last century. However, recent tide in the use of electronic appliances and requirement for short and mid-range wireless energy exchange has helped WPT getting tremendous attention, and studies on WPT systems has gained momentum in past few years \cite{10}. Modern day WPT techniques are mainly based on the mutual induction between two coils. Two fundamental principles, which are commonly used in such systems, are near-field non-radiative magneto-inductive effects and the resonant coupling between both the emitter and the receiver coils \cite{11,12,13,14}. Like resonant excitation (as discussed above), here also it is extremely important to maintain the resonance in both the coils. Otherwise it may result in decrease in the efficiency\cite{11}. Few studies has been put forward to solve this issue \cite{15,16}. WPT is also dependent on the coupling distance between the coils and it has been shown in few articles that in the strong coupling regime, it is possible to enhance efficiency for larger distances \cite{14}.   

In this work, we study WPT in a system of two inductively coupled coils by exploiting the adiabatic and the TQD techniques. Using coupled mode theory \cite{17} , we find out the governing equations (which are similar to the Schr\"odinger equation for two level system) and devise power transfer mechanism which is impervious to the coupling strength and distance between the coils and intrinsic losses present in the system.
  
The remainder of this article is organized as follows. In Sec.\ref{sec2}, at first we present a brief discussion on the TQD algorithm and a detailed study of the coupled LC circuits using the coupled mode theory,followed by application of adiabatic and TQD techniques in order to achieve wireless power transfer. Subsequently, in Sec.\ref{sec3}, we have discussed numerical calculations and relevant results followed by the conclusion of our work in Sec. \ref{sec4}.
\section{\label{sec2}Theoretical formulation}
\subsection{\label{A}Transitionless quantum driving}
 Shortcut techniques for adiabatic processes ensued from the limitations of the adiabatic theorem itself \cite{3}. Adiabatic theorem states that if a state $\ket{n(t = 0)}$ is an eigenstate of a Hamiltonian $H(t = 0)$ and the Hamiltonian evolves very slowly (adiabatically) with time, then $\ket{n(t)}$ will remain in its eigenstate throughout the evolution. However it is possible to inverse engineer the system Hamiltonian from a set of predefined eigenstates to drive the system beyond the adiabatic limit using TQD \cite{3}. According to this method, the transition probability between the instantaneous eigenstates could be made zero to achieve  exact evolution of the eigenstates. To understand it, let us consider the Hamiltonian:
\begin{equation}
H(t)= \sum_n \ket{n(t)}E_n(t)\bra{n(t)}
\label{Ham}
\end{equation}
The instantaneous eigenstates of this Hamiltonian are the adiabatic states, given by:
\begin{equation}
\ket{\Psi_n(t)}=e^{i\xi_n(t)}\ket{n(t)},
\label{eigen}
\end{equation}
where $\xi_n(t)$ is the adiabatic phase. We can define a unitary transformation to diagonalize $H(t = 0)$ in order to obtain the adiabatic Hamiltonian, as follows: 
\begin{equation}
H^{ad}(t)= U^{-1}H(t)U-i\hbar U^{-1}\dot{U}
\label{unitary}
\end{equation}
Here $U$ is a time dependent unitary operator. The first term on the RHS of Eq.~(\ref{unitary}), has  instantaneous eigenvalues and is diagonal itself. So it can drive the system along the adiabatic states $\ket{\Psi_n(t)}$. The second term refers to the non-adiabatic correction, and generally off-diagonal. When $H(t)$ changes slowly with time, the off-diagonal part of $H^{ad}(t)$ becomes negligible. Hence the probability for transfer among the instantaneous eigenstates tends to be zero but the evolution becomes slow. Problem arises when $H(t)$ changes fast with time, resulting in the breakdown of the adiabatic condition, as the off-diagonal part of $H^{ad}(t)$ becomes stronger and the system no longer follows the adiabatic path. In such a scenario, the TQD algorithm could be used to negate the effects of the non-adiabatic terms. According to this technique, we need to add an additional coupling Hamiltonian to cancel out the off-diagonal terms from $H^{ad}(t)$, so that it becomes diagonal once and for all and the system can be driven exactly regardless of the speed of the time evolution. In other words, one can drive the system infinitely fast along the adiabatic path. The additional Hamiltonian in the $\ket{n(t)}$ basis, is  given by:
\begin{equation}
H_1(t) = i\hbar\sum_m[\ket{\partial_t n(t)}\bra{n(t)}
- \braket{n(t)|\partial_t n(t)}\ket{n(t)}\bra{n(t)}]
\label{driving_ham}
\end{equation}
The total effective Hamiltonian, $H_{eff}(t) = H(t)+H_1(t)$, drives the system exactly beyond the adiabatic limit. One needs to configure $H_1(t)$ appropriately, depending on the system, to drive it in infinitely short time.
\subsection{\label{B}Coupled mode Theory for two interacting coils}
\begin{figure}
\includegraphics[height=6cm,width=8cm]{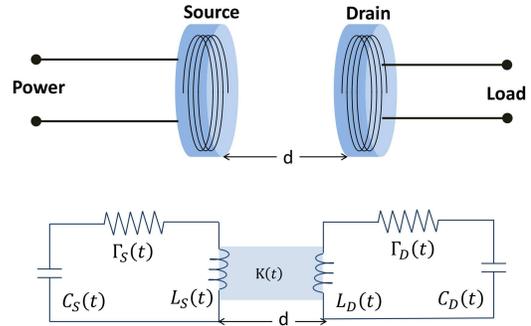}
\caption{\label{fig1} (Color online) (a) Typical wireless power transfer system consists of two coils separated by a distance $d$, (b) Schematic of the coils. Two lossy $LC$ circuits, {\it Source} and {\it Drain} with losses $\Gamma_s$ and $\Gamma_d$ respectively, coupled to each other by inductive coupling. The resonant frequencies are $\omega_s$ and $\omega_d$ and also $\omega_s \ne \omega_d$. }
\end{figure}
Let us consider a loss-less $LC$ circuit, without loss, with current $i(t)$ flowing in it and voltage, $v(t)$, across $L$ and $C$. The equations describing the voltage and the current can be written in terms of the following coupled differential equations:
\begin{subequations}\label{VI}
\begin{align}
v(t) = L\frac{di(t)}{dt},\\
i(t) = -C \frac{dv(t)}{dt}
\end{align}

\end{subequations}
Above equations could be rewritten, as a second order differential equation for voltage, as follows:  
\begin{equation}
\frac{d^2v(t)}{dt^2}+\omega_0^2 v(t) = 0,
\label{VV}
\end{equation}
where $\omega_0^2 = 1/LC$. The coupled equations in Eq.~(\ref{VI}) can be expressed by two uncoupled equations for mode amplitudes $a_+(t)$ and $a_-(t)$:
\begin{eqnarray}
&    &\frac{da_{\pm}(t)}{dt} =\pm j\omega_0 a_{\pm}(t),\\
\label{ampeqn}
\text{where    } \qquad &    &a_{\pm}(t) = \sqrt{\frac{C}{2}}(v(t) \pm j\sqrt{\frac{L}{C}}i(t))
\label{amps}
\end{eqnarray}
Here, $j=\sqrt{-1}$. The solutions for the current and the voltage from Eq.~(\ref{VI}) and Eq.~(\ref{VV}), when subjected to proper boundary conditions, are given by:
\begin{eqnarray}
&    &v(t)= |V| \cos (\omega_0 t),\\
\label{volt}
\text{and    } \qquad &    &i(t) = \sqrt{\frac{C}{L}}|V| \sin (\omega_0 t)
\label{cur}
\end{eqnarray}
Here $|V|$ is the peak voltage. Using above solutions, the total energy of the system could be obtained as: $ |a_{\pm}|^2 = \frac{C}{2} |V|^2 = W$, where $a_+$ being the positive frequency component of the mode amplitudes, while $a_-$ is its negative counterpart. In the rest of the analysis,we will consider the positive frequency component only and will drop the `+' subscript for simplicity. Taking loss in the system into account,  the equation is written in the following modified form: 
\begin{eqnarray}
\frac{da(t)}{dt} = j\omega_0 a(t) - \Gamma a(t),
\label{amp_loss}
\end{eqnarray}
where $\Gamma$ is the decay rate due to the dissipation from the coils. In our study we consider two such coils, namely the {\it Source} and the {\it Drain} (Fig.~(\ref{fig1})), which are coupled by mutual inductance between them. These two coils are off resonant to each other having different resonant frequencies $\omega_S$ and $\omega_D$. The coupling between the coils is given by $\kappa(t) = M(t)\sqrt{\omega_s\omega_d/L_s L_d}$, where $M(t)$ is the mutual inductance between these two coils. The mode amplitudes are $a_s(t)$ and $a_d(t)$ respectively and are coupled to each other. Total energy of the system is $|a_s(t)|^2 + |a_d(t)|^2$. This system can be expressed using the following set of coupled equations:
\begin{subequations}
\begin{align}
\frac{da_s(t)}{dt} = (j\omega_s - \Gamma_s) a_s(t) + j \kappa a_d(t)\\
\frac{da_d(t)}{dt} = (j\omega_d - \Gamma_d - \Gamma_w) a_s(t)+ j \kappa a_s(t)
\end{align}
\label{amp_loss}
\end{subequations}
Here $\Gamma_s$ and $\Gamma_d$ are the intrinsic loss rates of the {\it source} and the {\it drain} respectively. $\Gamma_w$ is the work extracted from the {\it drain} coil.
\subsection{\label{C}Energy transfer protocols}
In order to design the energy transfer protocols, we discuss two methods here: the adiabatic passage and transitionless quantum driving. From the coupled mode theory, discussed above, we can characterise the system by defining the Hamiltonian in $[a_s(t) \quad a_d(t)]^T$ basis, as follows:
\begin{equation}
H(t) = \begin{pmatrix} -\omega_s-j\Gamma_s & -\kappa(t) \\
 -\kappa(t) & -\omega_d-j\Gamma_d-j\Gamma_w \end{pmatrix}
\label{2level_ham}
\end{equation}
As the circuits,chosen here, are not resonant to each other, we consider the system in a rotating frame of reference, where the interaction Hamiltonian in the diabatic basis is as follows:
\begin{equation}
H(t) = \begin{pmatrix} \frac{\Delta(t)}{2}-j\Gamma_s & -\kappa(t) \\
 -\kappa(t) & -\frac{\Delta(t)}{2}-j\Gamma_d-j\Gamma_w \end{pmatrix}
\label{2level_INT}
\end{equation}
The diabatic basis are given by, $b_{s,d}(t)=a_{s,d}\exp[-j(\omega_s+\omega_d)t/2] $.In this new basis, one have to consider only the frequency difference between the coils, given by, $\Delta(t)=\omega_d(t)-\omega_s(t)$.
\subsubsection{\label{C1}Adiabatic following}
The instantaneous eigenvectors of Eq.~(\ref{2level_INT}) without losses are, 
%
\begin{subequations}
\begin{align}
B_+(t) &= \cos(\frac{\Theta(t)}{2}) b_s(t) - \sin({\frac{\Theta(t)}{2}}) b_d(t) \label{B+}\\
B_-(t) &= \sin(\frac{\Theta(t)}{2}) b_s(t) + \cos({\frac{\Theta(t)}{2}}) b_d(t) \label{B-}
\end{align}
\label{eigenvect}
\end{subequations}
Here $\Theta(t)$ is the angle of mixing,given by $\Theta(t) = \frac{1}{2}\tan^{-1}(2\kappa/\Delta)$. $B_{\pm}(t)$ are also known as adiabatic states. For adiabatic evolution, one needs to vary the Hamiltonian infinitely slowly or adiabatically so that the system always follows a particular state $B_+$ or $B_-$ during a complete cycle of the time evolution. To achieve this, the evolution ought to follow the adiabatic condition, which is obtained by comparing the non-adiabatic correction terms (Eq.~(\ref{unitary})) to the instantaneous eigen-energies. It is expressed as follows:
\begin{equation}
|\langle \partial_t B_{\pm}(t)|B_{\mp}(t)\rangle| \ll \sqrt{(4\kappa(t)^2+\Delta(t)^2)}
\label{adiabaticity}
\end{equation}
The fulfillment of this condition makes the transition probability between $B_+(t)$ and $B_-(t)$ zero. Moreover, if we assume that the system is initially in the state $B_+(t)$ and the power is in the {\it source} coil at $t=0$, which refers to $\Theta (0)=0$ , then,  by rotating $\Theta$  clockwise to $\frac{\pi}{2}$ , one could arrive at the final state, $b_d(t)$. Thus the power ends up in the {\it drain} coil. This rotation could be achieved by sweeping $\Delta(t)$ from a large negative value to a large positive value. For our system, we chose $\kappa(t)$ and $\Delta(t)$ according to the well known Landau-Zener scheme \cite{18}, given by:
\begin{equation}
\kappa (t) = \kappa_0 ;\qquad \Delta(t) = \delta + \beta(t-t_0)
\end{equation}
where $\delta$ is some arbitrary offset between the frequencies of the two coils and $\beta$ determines the slope in $\Delta(t)$, which eventually controls the speed of the process. Under such choices, large $t_0$ is needed to satisfy the adiabatic condition and  $t_0$ ,effectively, determines the width of the evolution cycle. When the loss rates are non-zero in both the {\it source} and the {\it drain} coil, the system could be considered as dissipative. These dissipations can be modelled mathematically via the dissipation matrix: 
\begin{equation}
\Gamma = \begin{pmatrix}  \Gamma_s & 0 \\
 0 & \Gamma_d+\Gamma_w \end{pmatrix}
\label{loss}
\end{equation}
It should be noted that, for adiabatic evolution, the intrinsic loss rates $\Gamma_s$ and $\Gamma_d$ should be less than the coupling strength $\kappa_0$, otherwise the evolution would not be possible and the power would be lost from the coil itself.
\begin{figure}
\includegraphics[height=6cm,width=9cm]{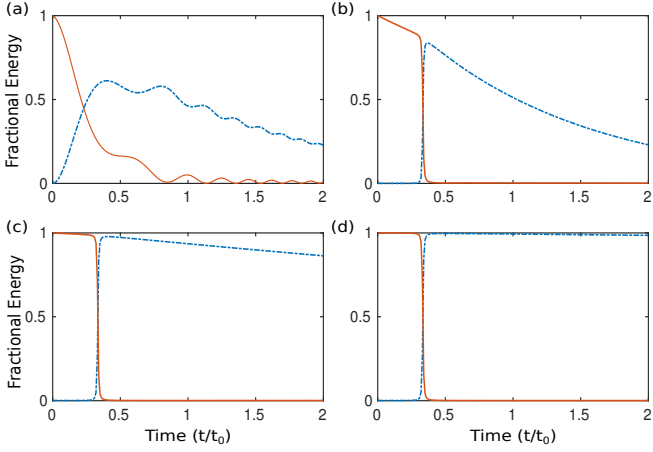}
\caption{\label{fig2} (Color online) Evolution of energy from the {\it Source} coil (solid red) to the {\it Drain} coil (dash-dotted blue) with $\Gamma_s=\Gamma_d=4\times 10^{3}s^{-1}$, $\delta=2\times 10^{5}s^{-1}$. (a) Adiabatic evolution for the time window $2t_0$ where $\kappa_0 = 4\times 10^{4}s^{-1}$, $\beta=3\times 10^{9}s^{-2}$ and $t_0 = 10^{-4}s$, followed by energy evolution using TQD with weaker coupling strenght $\kappa_0 = 4 \times 10^{2} s^{-1}$ and decreasing time windows (b) $\beta = 3\times 10^{9}s^{-2}$ and $t_0 = 10^{-4}s$, (c) $\beta = 3\times 10^{10}s^{-2}$ and $t_0 = 10^{-5}s$, (d) $\beta = 3\times 10^{11}s^{-2}$ and $t_0 = 10^{-6}s$, }
\end{figure}
\subsubsection{\label{C1}Shortcut to adiabaticity}
For shortcut, we first transform our Hamiltonian in Eq.~(\ref{2level_INT}), in the adiabatic basis, using Eq.~(\ref{unitary}),where the basis states are related as:
\begin{equation}
[B_+(t),B_-(t)]^T = U(\Theta(t))^\dagger[b_s(t), b_d(t)]^T
\label{adia_basis}
\end{equation}
The non-adiabatic correction terms are generally negligible under adiabatic approximation. The Landau-Zener formula for the total transition probability between adiabatic states in our case is, $p = \exp [-2 \pi \kappa_0^2/\beta]$. One can obtain $p \simeq 0$ when Eq.~(\ref{adiabaticity}) is satisfied, which can be written simply as $\beta \ll 8\kappa_0^2$. However if one wants to drive the evolution faster, non-adiabatic corrections becomes stronger and $p$ no longer remains zero. To avoid such a scenario, the adiabatic Hamiltonian needs to be diagonalized exactly. To serve this cause, we add the additional interaction to the adiabatic Hamiltonian as proposed by Berry \cite{3}.We use Eq.~(\ref{driving_ham}) to find the additional interaction. The second term in Eq.~(\ref{driving_ham}) vanishes owing to the orthogonality of $B_{\pm}(t)$. The first term can be written in the adiabatic basis as follows:
\begin{figure}
\includegraphics[height=5cm,width=8cm]{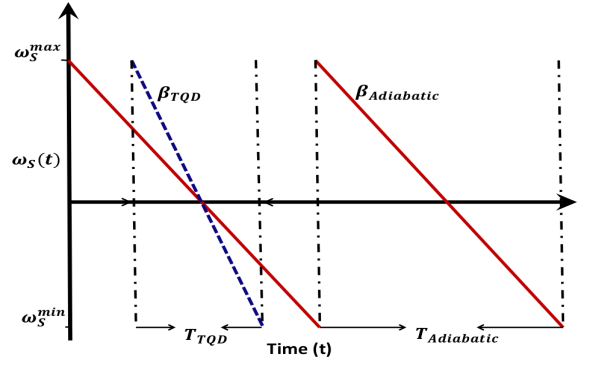}
\caption{\label{fig3}(Color online) Frequency sweep of $\omega_s(t)$ (or $\Delta(t)$ when $\omega_d =$ constant). The sweep is linear with slope $\beta$ according to L-Z model. For adiabatic evolution $|\beta| = \beta_{adiabatic}$ is small (solid red) and $|\beta| = \beta_{TQD}$ is high for the TQD method. Time period required for adiabatic case is large accordingly i.e.\ $T_{TQD}<T_{Adiabatic}$.}
\end{figure}
\begin{equation}
H_1(t) = j\begin{pmatrix}  0 & \kappa_a(t) \\
  -\kappa_a(t) & 0 \end{pmatrix}
\label{drive}
\end{equation} 
The total Hamiltonian, required for the transitionless driving, is given by
\begin{equation}
H(t) + H_{CD}(t) =\begin{pmatrix}  (\Delta-\dot{\phi})/2 & \sqrt{\kappa^2(t)+\kappa_a^2(t)} \\ 
\sqrt{\kappa^2(t)+\kappa_a^2(t)} & -(\Delta-\dot{\phi}/2) \end{pmatrix},
\label{transitionless}
\end{equation} 
where
\begin{equation}
\kappa_a(t) = \frac{|\dot{\Delta}(t)\kappa(t)-\dot{\kappa}(t)\Delta(t)|}{\Delta^2(t)+4\kappa^2(t)}
\label{theta}
\end{equation}
and
\begin{equation}
\dot{\phi}(t) = \frac{2\Delta(t)\dot{\Delta}^2(t)}{\Delta^2(t)+4\kappa^2(t)+\dot{\Delta}^2(t)}
\label{theta}
\end{equation}
Here $(\Delta-\dot{\phi})/2$ is the effective detuning and $\sqrt{\kappa^2(t)+\kappa_a^2(t)} = \kappa_{eff}$ is the modified coupling between the two coils. $\phi$ characterizes another unitary rotation given by 
\begin{equation}
\begin{pmatrix} b_s'\\b_d'\end{pmatrix} = \begin{pmatrix}  e^{-j\phi/2} & 0 \\
  0 & e^{j\phi/2} \end{pmatrix} \begin{pmatrix} b_s\\b_d\end{pmatrix}
\label{drive}
\end{equation} 
With all these unitary transformations, one needs to keep track of all the bases used in the process and keep them consistent. The mixing angle $\Theta$ should be adjusted properly via the boundary conditions, given by $\Theta(0) = 0$, $\Theta(T) = \pi/2$ and $\kappa_a(0)=\kappa_a(T) = 0$.
\section{\label{sec3}results and discussion}
To envisage the transfer of energy from the {\it source} to the {\it drain} ,with the effects of intrinsic losses taken into account, we followed the standard density matrix approach and therefore solved the master equation, given by
\begin{equation}
\frac{d\rho}{dt} = -j[H,\rho] - \frac{1}{2} \{ \Gamma,\rho \}
\label{master}
\end{equation}
Here $\rho(t)$ is the density matrix and $\Gamma$ represents the dissipation matrix as described in Eq.~(\ref{loss}). The diagonal elements of $\rho(t)$ are $\rho_{ss}=|b_s|^2$ and $\rho_{dd}=|b_d|^2$ which represents the energy of the {\it source} and
\begin{figure}
\includegraphics[height=8cm,width=8cm]{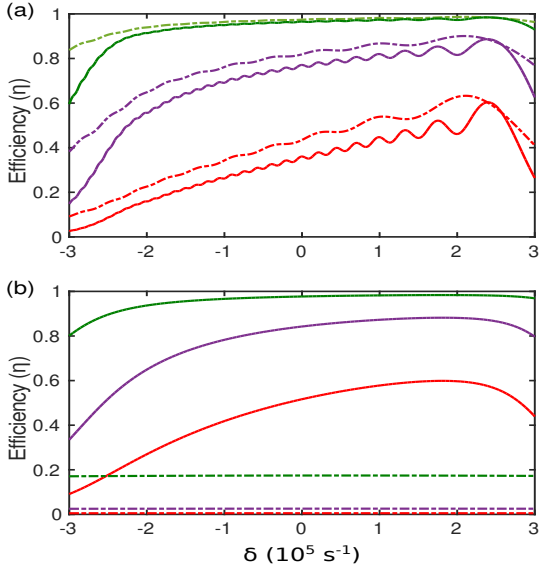}
\caption{\label{fig4}(Color online) Comparison of efficiency ($\eta$) as a function of $\delta$ between adiabatic (dashed-dotted) and tqd based methods (solid) for different $\kappa_0 / \Gamma_{s,d}$ values: $\kappa_0 / \Gamma_{s,d}=10$ (red), $\kappa_0 / \Gamma_{s,d}=50$ (purple), $\kappa_0 / \Gamma_{s,d}$=100 (green) where $\Gamma_w=10^4 s^{-1}$. Time windows ($T$) for the evolution are as follows: (a) $T=200 \mu s$, (b) $T=2 \mu s$.}
\end{figure}
 the {\it drain} coil respectively.   For adiabatic evolution we chose the interaction Hamiltonian in Eq.~(\ref{2level_INT}) and for TQD,  $H(t)$ is taken from Eq.~(\ref{transitionless}).

In Fig.~(\ref{fig2}), we present the results showing the evolution of fractional energies, $|b_{s,d}(t)|^2/|b_{s}(0)|^2$ , in presence of intrinsic losses, using both the adiabatic and the TQD algorithm. Even when the adiabatic condition is being satisfied, as shown in Fig.~(\ref{fig2}a) , as a result of adiabatic evolution, the fractional energy attains the value, on the order of $0.25$ or so at the end of the given time window. The requirement of large time, i.e.\ $t_0 = 10^{-4}s$, so that the adiabaticity condition is maintained, results in energy dissipation due to the intrinsic losses.

\begin{figure}
\includegraphics[height=8cm,width=8cm]{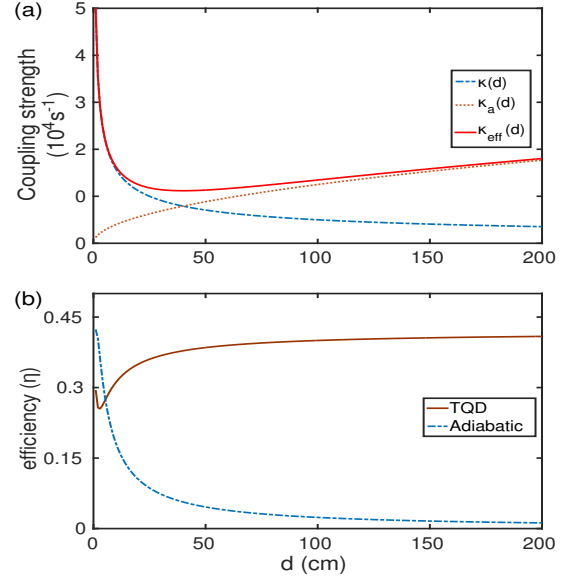}
\caption{\label{fig5}(Color online) Dependence on the distance $d$ of the (a) coupling $\kappa(d)$ between the {\it source} and the {\it drain} coil (dash-dotted blue), additional coupling $\kappa_a(d)$ (dotted brown) and effective coupling $\kappa_{eff}(d)$ for TQD (solid red), (b) efficiency $\eta(d)$ for adiabatic method (dash-dotted blue) and for TQD (solid brown) .}
\end{figure}
However, when TQD is applied, fractional energy of the {\it drain} coil almost attains nearly the same value for the same period of time (Fig.~(\ref{fig2}b)). But when the adiabatic condition is violated i.e.\ $\beta \geq 8\kappa_0^2$, enhancement of energy in the drain coil could be observed in Fig.~(\ref{fig2}c). The affects of intrinsic losses are almost eliminated as the time period becomes shorter and shorter(Fig.~(\ref{fig2}d)). It is worthwhile to note the effect of $\beta$ on the power transfer mechanism. Physically, $\beta$ determines the slope of $\Delta$ and thereby it controls the time period required to complete a single power transfer cycle. In the adiabatic regime, $\beta$ is relatively small and hence the required period $T_{adiabatic}$ is larger and because of that, power is dissipated from the source coil during the process. When $\beta$ is large, the frequency sweep becomes faster and $\omega_s(t)$ becomes steeper (here $\omega_d$ is taken as constant) as shown in Fig.~(\ref{fig3}) which results in squeezing of the period. Thus the power is transferred to the {\it drain} in short time with minimum loss from the {\it source}. Also it is obvious that one needs to repeat the cycle over and over again to transfer power for practical purposes.   

The work efficiency of our system is defined as the ratio between the useful extracted power from the drain, $P_{work}=\Gamma_w \int_0^T|b_d(t)|^2 dt$ to the total time averaged power $P_{total}$ in the system over a particular time period $T$, given by
\begin{equation}
\eta = \frac{\Gamma_w \int_0^T|b_d(t)|^2 dt}{\Gamma_s \int_0^T|b_s(t)|^2 dt +(\Gamma_d+\Gamma_w) \int_0^T|b_d(t)|^2 dt}
\label{eff}
\end{equation}
\begin{figure*}
\includegraphics[height=7cm,width=15cm]{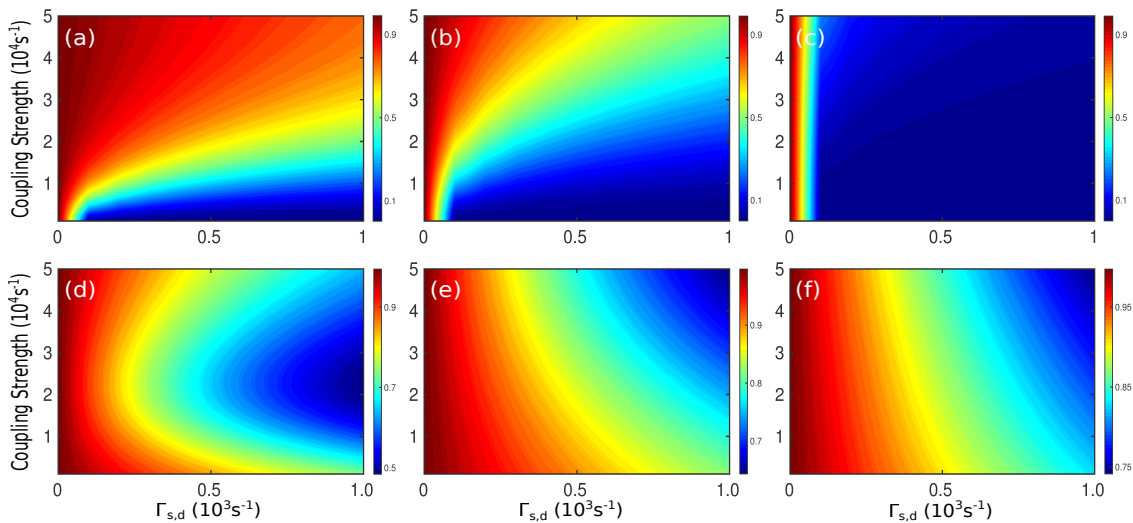}
\caption{\label{fig6} (Color online) Contour plots for efficiency with respect to the variations in $\kappa$ ($\times 10^{4}s^{-1}$) and intrinsic losses $\Gamma_s= \Gamma_d$ ($\times 10^{3}s^{-1}$) and $\Gamma_w=10^4s^{-1}$. (a), (b), (c) for adiabatic case and (d), (e), (f) for TQD method with $t_0 = 10^{-4}s$ in (a) and (d), $t_0 = 10^{-5}s$ in (b) and (e) and $t_0 = 10^{-6}s$ in (c) and (f)}
\end{figure*}
We studied efficiency of the system against the variation of the initial frequency difference between the coils for different values of $\kappa_0/\Gamma_{s,d}$. The efficiency strongly depends on the coupling strength $\kappa_0$ for both the adiabatic and the TQD approach and it increases with increasing $\kappa_0$ as depicted in Fig.~(\ref{fig4}a) . Although this is not that surprising, but in the beyond adiabatic regime (for shorter periods), efficiency for adiabatic method decreases rapidly. However efficiency remains intact for TQD algorithm which can be seen from Fig.~\ref{fig4}b. 

The coupling between the coils for WPT systems are generally very sensitive to the distance, $d$ , between the coils and it decays very rapidly with larger coupling distances \cite{11,12}. In Fig.~\ref{fig5}(a) and \ref{fig5}(b) we depict the dependence of coupling strength $\kappa(d)$ and efficiency $\eta(d)$ respectively on the distance $d$ between the coils. From Fig.~(\ref{fig5}b) it is clear that, even in the adiabatic regime, the efficiency for the adiabatic case decreases with increasing $d$. $\eta$ tends to zero for $d=2m$ or so as the strength of $\kappa$ goes down. But in the case of TQD, $\eta$ maintains a steady value for large $d$ and that certainly gets enhanced with larger $\beta$ values. The reason behind such behavior could be understood from the formalism of inverse engineering. We can observe from  Fig.~(\ref{fig5}a) that as $\kappa$ decreases with distance, we require an additional coupling $\kappa_a$ which increases with distance so that the effective coupling $\kappa_{eff}$ constitutes a reasonable strength for sustained power transfer over a certain range of distance.

We have also studied the efficiency of our scheme against the variations in $\kappa_0$ and the intrinsic losses ($\Gamma_s=\Gamma_d$). From the contour plots in Fig.~(\ref{fig6}), we observe that $\eta$ is nearly unity for lower losses and higher $\kappa_0$ in the adiabatic regime. As we move from Fig.~(\ref{fig6}a) to Fig.~(\ref{fig6}c), adiabaticity gradually breaks down and efficiency also decreases gradually. Finally it goes to zero when $t_0 = 10^{-6}$s for any reasonable amount of losses. On the other hand, using TQD, we find that the achievable efficiency is highly robust against variations in $\kappa_0$ and $\Gamma_{s,d}$ unlike its adiabatic counterpart. Also the efficiency is found to get enhanced with decreasing time window.
 
\section{\label{sec4}Conclusion}
In conclusion, we have explored a WPT system in the light of adiabatic method and transitionless quantum driving method. Our findings could be summarized as follows. The adiabatic evolution of power is a useful way to transfer power between two coils. Unlike resonant WPT systems, it uses two off-resonant coils and frequency sweeping is used for power transfer. However, it has to fulfil the adiabatic condition which makes the time required to transfer power in each cycle longer, resulting in dissipation of power from the source coil and hence efficiency decreases. On the other hand, the TQD algorithm enables us to enhance the efficiency of power transfer. The algorithm suggest that, it is possible to decrease the transfer time and increase the efficiency by invoking an additional interaction between the coils.The amount of power dissipated from the {\it source} also gets decreased. The WPT method using TQD shows more robustness compared to the adiabatic one against the variations in the parameters like the coupling strength, intrinsic losses and the coupling distance.
\\

\setcounter{secnumdepth}{0}
\section{ACKNOWLEDGMENTS}
K. P. would like to gratefully acknowledge the research fellowship provided by MHRD, Govt. of India.

\end{document}